\documentclass[10pt]{iopart}
\pdfoutput=1 
\usepackage{siunitx}             
\usepackage{graphicx}            
\usepackage{xcolor}              
\definecolor{njplinkcolor}{RGB}{24, 131, 184} 
\usepackage[colorlinks,linkcolor={njplinkcolor},citecolor={njplinkcolor},urlcolor={njplinkcolor}]{hyperref} 

\hypersetup{pdfauthor={Jo Rushton, Ritayan Roy, James Bateman and Matt Himsworth}, pdftitle={A dynamic magneto-optical trap for atom chips}, pdfkeywords={magneto-optical trap, laser cooling, cold atoms}}
\DeclareSIQualifier\peak{p}      
\DeclareSIQualifier\peakpeak{pp} 
\DeclareSIUnit\sample{S}
\DeclareSIUnit\gauss{G}          

\begin{document}

\title{A dynamic magneto-optical trap for atom chips}

\author{Jo Rushton$^1$, Ritayan Roy, James Bateman$^\dagger$ and Matt Himsworth$^2$}

\address{School of Physics and Astronomy, University of Southampton, Highfield, Southampton, SO17 1BJ, UK}
\address{$^\dagger$Present address: Department of Physics, College of Science, Swansea University, Swansea, SA2 8PP, UK}
\eads{$^1$\mailto{physics@josephrushton.com}, $^2$\mailto{m.d.himsworth@soton.ac.uk}}
\vspace{10pt}

\begin{abstract} 
We describe a dynamic magneto-optical trap (MOT) suitable for the use with vacuum systems in which optical access is limited to a single window. This technique facilitates the long-standing desire of producing integrated atom chips, many of which are likely to have severely restricted optical access compared with conventional vacuum chambers. This `switching-MOT' relies on the synchronized pulsing of optical and magnetic fields at audio frequencies. The trap's beam geometry is obtained using a planar mirror surface, and does not require a patterned substrate or bulky optics inside the vacuum chamber.  Central to the design is a novel magnetic field geometry that requires no external quadrupole or bias coils which leads toward a very compact system. We have implemented the trap for $^{85}$Rb and shown that it is capable of capturing 2 million atoms and directly cooling below the Doppler temperature.
\end{abstract}

\pacs{37.10.De, 37.10.Gh}
%
\vspace{2pc}
\noindent{\it Keywords}: magneto-optical trap, laser cooling, cold atoms
%
%
%
%

\section{Introduction} 

The magneto-optical trap (MOT) has revolutionized the fields of atomic and quantum physics by providing a gateway between the thermal `classical' regime down to the ultracold `quantum' regime where the de-Broglie wavelength becomes significant and environmental decoherence is greatly reduced. Dense samples of ultracold atoms can then be manipulated with exquisite detail by optical and magnetic fields for a variety of fundamental and applied tasks. An `atom chip' is an arrangement of microfabricated current-carrying wires patterned on a substrate which is used to trap and control atoms via the strong magnetic field gradients offered at distances close to conductors \cite{schmiedmayer1995guiding, weinstein1995microscopic, fortagh1998miniaturized, denschlag1999guiding, reichel2011atom}. Atoms chips enable highly sophisticated experiments to be condensed into areas on the order of a few square centimetres and readily lend themselves to the miniaturization and integration of cold atom systems for practical applications beyond the laboratory \cite{salim2011compact}. 

The beam geometry of the standard six-beam MOT is unsuitable for operation near to all but transparent substrates and so needed to be adapted for use with atom chips. One of the most notable variations of the trap is the mirror-MOT (M-MOT), which has been an indispensable tool in loading atom chips since their first demonstration \cite{ReichelMirrorMOT, folman2000controlling}. The trap's popularity is not only due to its ability to hold atoms close to surfaces, but also because of its simplicity, requiring only a pair of anti-Helmholtz coils and a mirror.  More recently efforts have begun to be directed towards portable systems, including so called integrated atom chips, where the ubiquitous multi-window stainless steel UHV chamber is replaced by microfabricated vacuum cells, and tables of optics and electronics are miniaturized into portable packages.  These devices, such as the in development micro-MOT \cite{IntegratedAtomChipFeasibility}, promise to make cold atom technology practical outside of the laboratory, but require our existing manipulation techniques to adapt to new constraints, in particular a reduction in optical access. 

The geometry of the standard M-MOT consists of an anti-Helmholtz coil tilted at \SI{45}{\degree} to the surface of a mirror and positioned such that its magnetic field zero is just above the mirror's surface. A counter-propagating beam pair is directed  towards the magnetic field zero along the axis of the coils and a second counter-propagating beam pair is aligned parallel to the mirror's surface, perpendicularly to the other beams and intersecting them at the field minimum.  These latter, unreflected, beams are those that are problematic if the optical access is restricted to a single window. Additional optics may be placed inside the chamber to redirect these beams but in order to trap a significant number of atoms the chamber volume needs to be increased accordingly, which runs contrary to our aim of miniaturizing the system.  The pyramidal, tetrahedral, and grating MOTs \cite{Lee96PyramidMOT, SiPyramidMOT, TetraMOT, GMOT} alleviate this issue of optical access, each of which only require a single beam to operate. The drawback to these devices is that they have complex and expensive microfabrication procedures and several of these designs are not easily compatible with planar atom chip structures.

Here we demonstrate a variation of the M-MOT which only requires optical access through a single viewport, is able to capture a modest number of atoms, and can cool below the Doppler temperature without an additional sub-Doppler stage.  Our design has inherently low scatter, can be used to trap multiple atomic species simultaneously and is well suited for use in integrated atom chips, where optical access is restricted.  This new design is a time varying trap in a similar vein to the AC-MOT \cite{ACMurrayMOT}, and as a result is named the switching-MOT (S-MOT).

\section{Theory}
\begin{figure}
	\centering
	\includegraphics[width=\textwidth]{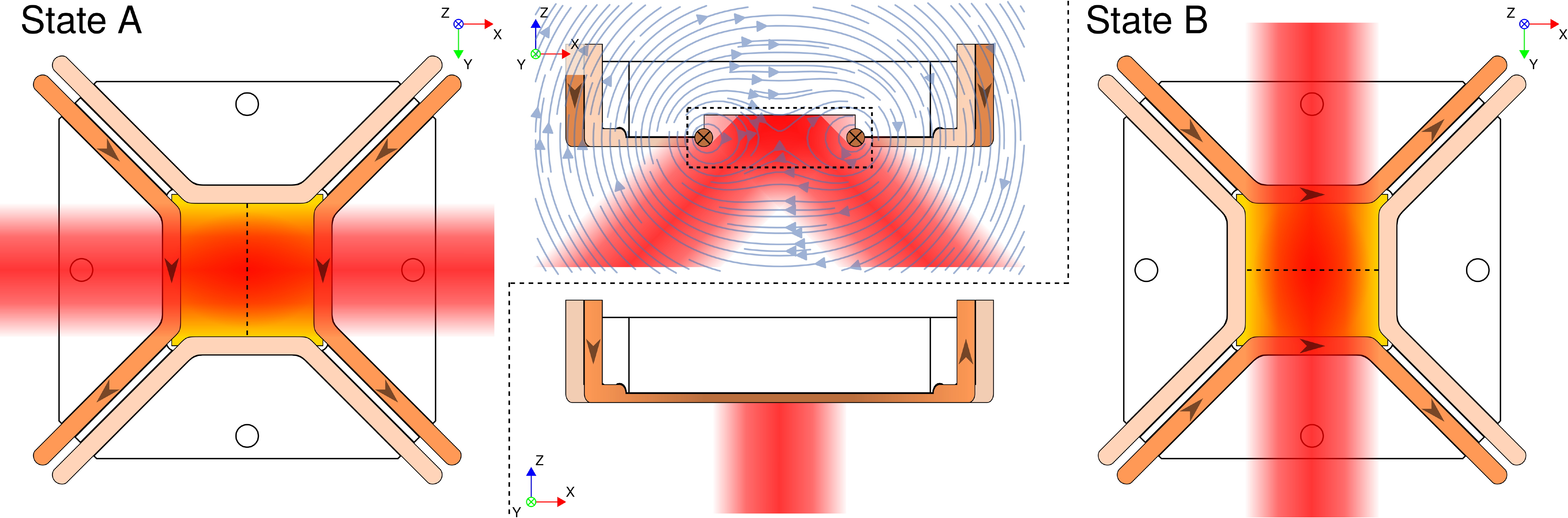} 
	\caption{The two states between which the S-MOT alternates.  The highlighted wires in each state corresponds to the active pair which carry current in the same direction. The blue streamline plot shows the resulting 2D quadrupole magnetic field, which in combination with the associated counter-propagating beam pair causes 2D trapping at the midpoint of the active wire pair (dashed lines).}
	\label{smot_wires_and_field}
\end{figure}

The S-MOT arose from an attempt to modify the M-MOT so that all four beams are incident at an angle of \SI{45}{\degree} to the mirror, rather than two as in the standard design, thus eliminating the restrictive beams parallel to the mirror surface. In this geometry one counter-propagating beam pair is on a plane orthogonal to the other beam pair, but with an angle of \SI{60}{\degree} between neighbouring beams. To ease microfabrication we desired a planar wire geometry to generate the required quadrupole magnetic field for trapping. Earlier groups have removed the need for anti-Helmholtz coils in order to produce their quadrupole field, obtaining it instead by combining a bias field and that due to a current carrying wire \cite{wildermuth2004optimized}.  This method, however, still relies on external coils to generate the bias field and furthermore uses the conventional M-MOT beam geometry. We took a similar approach but without using any coils observing that, as shown in figure \ref{smot_wires_and_field}, the magnetic field could be emulated by a parallel pair of wires that both carry current in the same direction. This, however, is only a two-dimensional approximation to a quadrupole magnetic field and so can only confine atoms to a line equidistant between the wires. 

Three-dimensional trapping is not possible simply by introducing a second, orthogonal, wire pair to the design as the total field due to the sum of both wire pairs produces a line of zero magnetic field diagonal to the wires.  Three-dimensional confinement is possible with this geometry, however, if the trap is \emph{dynamic}: alternating between two states in an analogous scheme to the quadrupole ion trap.

Figure \ref{smot_wires_and_field} illustrates the S-MOT's geometry.  At any moment in time current only passes through one wire pair, whilst in the same instance only one counter-propagating beam pair is directed to the magnetic field zero at an angle of \SI{45}{\degree} to the mirror. Each of the S-MOT's states provide a trapping force towards either of the lines $x=z=0$ or $y=z=0$, represented on figure \ref{smot_wires_and_field} as dashed strokes.  Rapidly switching between these states causes the atoms to see the average of these forces, coalescing them at the origin located at the centre of the two wire pairs.

\subsection{The need for optical switching}\label{subsection:The need for Optical Switching}
It should be apparent from the earlier discussion why the magnetic fields need to be switched in order to produce trapping in the S-MOT, however the need for the optical switching is not so obvious.

To understand where this requirement originates we must first realize that in the regime of low intensity light, atoms that are stationary in a MOT experience a force from a counter-propagating pair of beams with wavevectors $\pm\vec{k}$ which is proportional to $\pm(\vec{k}\cdot\vec{B})\hat{k}$, where $\vec{B}$ is the magnetic field at the atom's location and the sign is determined by the choice of circular beam polarization \cite{JRushtonThesis}.  From this we can understand that in order to trap atoms within a MOT we must ensure that $\pm(\vec{k}\cdot\vec{B})\hat{k}$ always points towards the trap centre, or in the case of the S-MOT, that for the beams active in each time-step this expression always points towards the corresponding centre line of zero magnetic field.

\begin{figure}
	\centering
	\includegraphics[width=\textwidth]{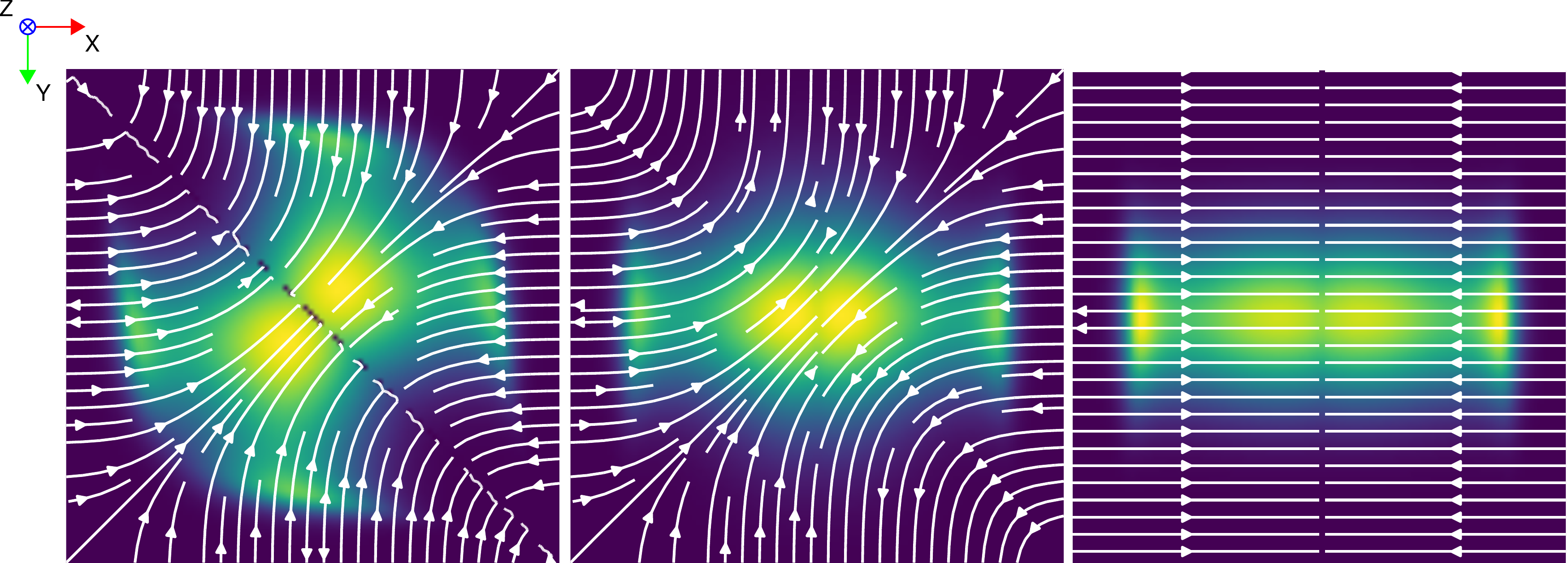}
	\caption{Streamline and density plots of the force upon a stationary atom in the $xy$-plane of the S-MOT. \emph{Left:} shows the result of the S-MOT operating in the DC configuration, with the lasers in both axes unshuttered and the wires constantly passing current.  \emph{Middle:} is the result of the S-MOT operating with unshuttered lasers but with the wires in AC mode (shown in state A).  \emph{Right:} shows the force plot of the S-MOT with both the magnetic fields and lasers being switched (shown in state A). The left two streamline plots may appear to be identical, however the line of minimum force is along $y=x$ for the left plot and $x=0$ for the middle plot.}
	\label{SMOT_force_figure}
\end{figure}

In order to make the S-MOT's behaviour easier to interpret, in the following discussion the laser beams are treated as being infinitely large and of uniform intensity.  Here we consider the magnetic field due a pair of wires that lie in the $xy$-plane with the current in the $+y$ direction and the trapping region at the origin. We only need to analyse the trapping force in a single time step, as when the wires are switched the situation is identical and is merely rotated \SI{90}{\degree} around the $z$ axis.  In this scheme the cooling beams are directed to the line of magnetic field zero at a \SI{45}{\degree} angle of incidence, corresponding to the wavevectors $\vec{k}_{xz1}=|\vec{k}|(\hat{x} + \hat{z})/\sqrt{2}$ and $\vec{k}_{xz2}=|\vec{k}|(\hat{x} - \hat{z})/\sqrt{2}$. These describe a beam in the $xz$-plane before and after reflection in the S-MOT's mirror, whilst the two retro-reflected beams are not written explicitly.   The magnetic field generated by the S-MOT's wires can be roughly approximated by:
\begin{eqnarray}
\vec{B}=G(z\hat{x} + x\hat{z})\,,
\end{eqnarray}
where $G$ is the gradient of the field. The force exerted by a counter-propagating pair of beams with wavevectors $\pm\vec{k}_{xz1}$ is
\begin{eqnarray}
\vec{F}_{xz1}\propto\pm(\vec{k}_{xz1}\cdot\vec{B})\hat{k}_{xz1}&=\pm\frac{|\vec{k}|}{\sqrt{2}}(\hat{x} + \hat{z})\cdot G(z\hat{x} + x\hat{z})\hat{k}_{xz1} \nonumber \\
&=\pm\frac{G}{\sqrt{2}}(z+x)\vec{k}_{xz1}\,.
\end{eqnarray}
Similarly the force due to the counter-propagating beams with wavevectors $\pm\vec{k}_{xz2}$ is
\begin{eqnarray}
\vec{F}_{xz2}\propto\mp(\vec{k}_{xz2}\cdot\vec{B})\hat{k}_{xz2}=&\mp\frac{G}{\sqrt{2}}(z-x)\vec{k}_{xz2}\,,
\end{eqnarray}
where the signs are opposite to those of $\vec{F}_{xz1}$ due to the beams having opposite helicities as a result of the reflection on the S-MOT's mirror.  The total force due to all of the beams in the $xz$-plane is
\begin{eqnarray}
\vec{F}_{xz}=\vec{F}_{xz1}+\vec{F}_{xz2}\propto&\pm G|\vec{k}|(\vec{x}+\vec{z})\,,
\end{eqnarray}
which with the correct choice of helicity is a restoring force in the $xz$-plane, and has no influence along the $y$ axis.  If we now consider the presence of a set of beams in the $yz$-plane corresponding to the wavevectors $\vec{k}_{yz1}=|\vec{k}|(\hat{y} + \hat{z})/\sqrt{2}$ and $\vec{k}_{yz2}=|\vec{k}|(\hat{y} - \hat{z})/\sqrt{2}$ (again the counter-propagating beams are not written explicitly.), then the force due a pair of counter-propagating beams with wavevectors $\pm\vec{k}_{yz1}$ is
\begin{eqnarray}
\vec{F}_{yz1}\propto\pm(\vec{k}_{yz1}\cdot\vec{B})\hat{k}_{yz1}=&\pm\frac{G}{\sqrt{2}}x\vec{k}_{yz1}\,.
\end{eqnarray}
Similarly the force due to the counter-propagating beams with wavevectors $\pm\vec{k}_{yz2}$ is
\begin{eqnarray}
\vec{F}_{yz2}\propto\mp(\vec{k}_{yz2}\cdot\vec{B})\hat{k}_{yz2}=&\pm\frac{G}{\sqrt{2}}x\vec{k}_{yz2}\,,
\end{eqnarray}
where once again the sign was flipped as a result of the change in helicity of the light upon reflection.  The total force due to all of the beams in the $yz$-plane is
\begin{eqnarray}
\vec{F}_{yz}=\vec{F}_{yz1}+\vec{F}_{yz2}\propto&\pm G|\vec{k}|x\hat{y}\,.
\label{eq:SMOTantitrapping}
\end{eqnarray}
This force provides a degree of anti-trapping regardless of the choice of helicity of the beams, as shown in figure \ref{SMOT_force_figure}, and justifies the need for them to be shuttered in synchronization with the magnetic field switching.

\section{Experiment}
\begin{figure}
\includegraphics[width=1.0\textwidth]{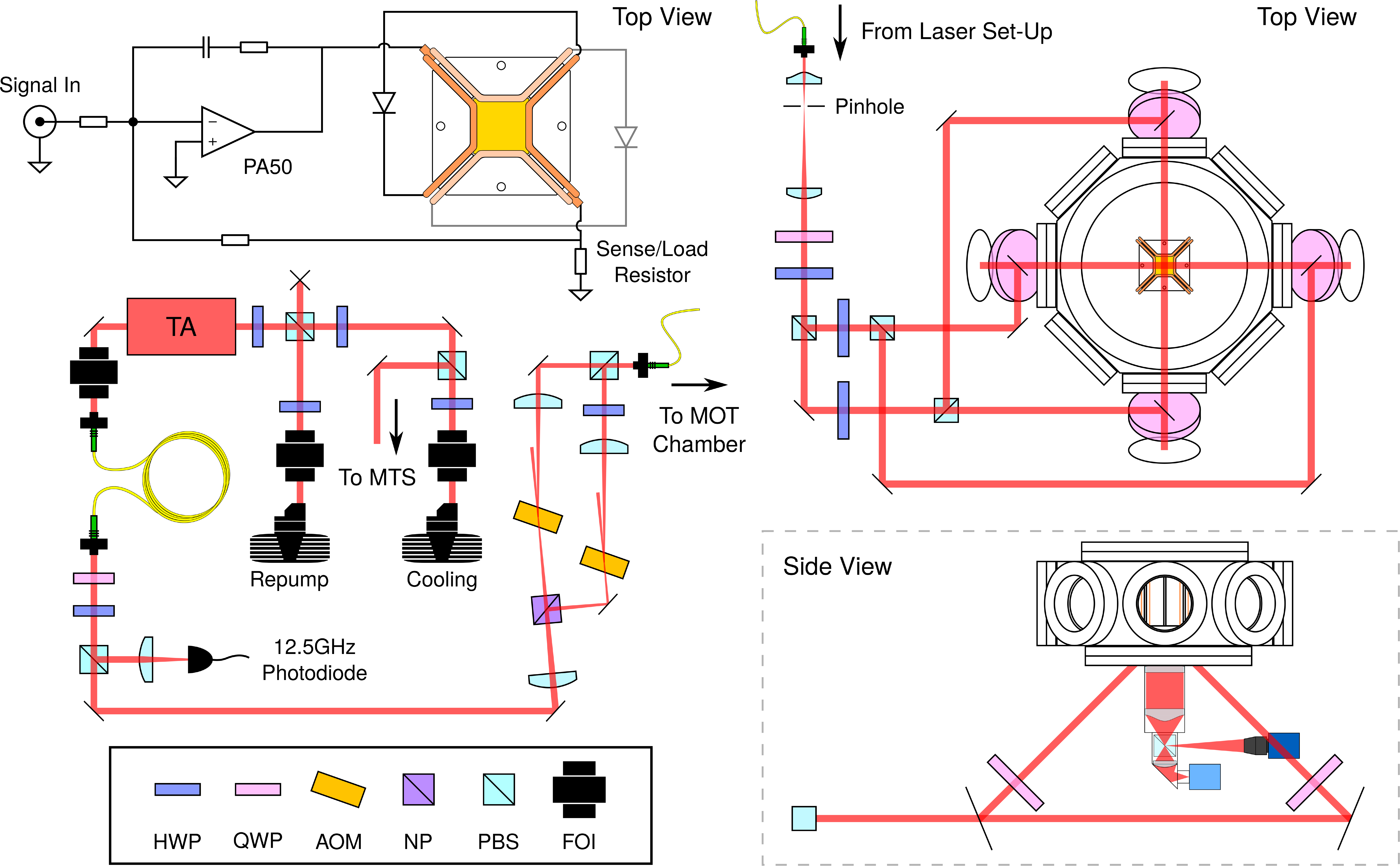}
\caption{Schematic of the experimental set-up used to generate the cooling and repump light and to perform their shuttering.  \emph{Top left:} a simplification of the circuit attached to the S-MOT's wires that, in combination with an input AC waveform, produces the correct current sequence for trapping with the S-MOT. \emph{Bottom left:} the laser systems, and optical switching elements. \emph{Right:} the beams are cleaned and separated before being directed into the vacuum chamber, and where the imaging optics are located. }
\label{experimental_setup}
\end{figure}

The laser set-up for the experiment is shown in figure \ref{experimental_setup}. A home-made external cavity diode laser provides the cooling light, which is frequency stabilized to the $5^{2}\text{S}_{1/2}, F=3 \rightarrow{} 5^{2}\text{P}_{3/2}, F'=4$ cooling transition of ${^{85}}$Rb via modulation transfer spectroscopy (MTS) \cite{Shirley:82}.  The beam from this laser is combined with that from a similar, repump, laser and is used to seed a m2k TA-0785-2000-DHP tapered amplifier (TA), whose output is cleaned by passing through a single-mode fibre. A portion of the TA's output is directed onto a fast photodiode (EOT ET-4000) and the resulting beat note is used to offset lock the repump from the cooling laser by the method described in \cite{PhaseLock}.

To perform the optical switching required for the S-MOT, a pair of \SI{80}{\mega\hertz} acousto-optic modulators (AOMs) are employed as shutters.  As shown in figure \ref{experimental_setup}, the beam combining both cooling and repump frequencies is divided into vertical and horizontal polarized components, each of which passes separately through one of the two shuttering AOMs. In addition to blanking the beams, these AOMs also impart an undesirable shift in the frequency of the repump and cooling light equal to the AOMs' driving frequency. In order to compensate for this frequency shift the MTS pump beam is obtained by double-passing the light from the cooling laser through a third \SI{80}{\mega\hertz} AOM.  As a result of this arrangement the MTS acts as an offset lock, stabilizing the light emerging from the laser to a point separated from the cooling transition by the MTS AOM's driving frequency \cite{JRushtonThesis,Negnevitsky:13,NegnevitskyThesis}.  The post-shuttered light is hence detuned from the cooling transition by the difference in the driving frequencies between the shuttering and MTS AOMs, and this small difference is used to set the red-detuning necessary for cooling and trapping.

After passing through the shutters the orthogonally polarized beams are recombined and then coupled into a non-polarization-maintaining optical fibre leading to the MOT chamber. At the other end of the fibre the beams are thoroughly cleaned with a spatial filter before being expanded to a $1/e^2$ radius of \SI{3.95}{\milli\metre}. The individually shuttered beams are then separated with a polarizing beam splitter and directed into the vacuum chamber. The chamber has a single anti-reflection coated viewport and is maintained at a pressure of \SI{2e-9}{\milli\bar} as measured from the lifetime of its trapped atoms \cite{ArpornthipPressure}. The geometry of the beams and lack of quadrupole magnetic field coils provides excellent optical access for detection, which we take advantage of by obtaining an NA of $\sim$0.6 using a Thorlabs ACL5040U-B aspheric condenser lens. The Earth's magnetic field is reduced with 3 external nulling coils, however we discuss in section \ref{Discussions} how an adapted version of the trap simplifies the nulling field geometry.   
  
In order to produce the sequence of current pulses required by the S-MOT, a circuit consisting of a pair of anti-parallel diodes was connected to the S-MOT's wires as shown in  figure \ref{experimental_setup}.  This circuit acts as a pair of half wave rectifiers when driven by an AC waveform, alternating the current through each of the wire pairs every half of its period. This circuit is placed in the feedback loop of a home-made current amplifier, which during switching delivers \SI{20}{\ampere\peak} half-rectified sine waves to the SMOT's wires at frequencies up to \SI{60}{\kilo\hertz}. Each of the SMOT's quadrupole wire pairs are composed of \SI{2.39}{\milli\metre} diameter wire separated by \SI{20}{\milli\metre} which, according to our simulations, produce a peak magnetic field gradient of \SI{6.5}{\gauss\per\centi\metre} along the direction of the beams.  In order to ameliorate induction related effects at these frequencies we avoided the use of coils in our design, used litz wire to connect the amplifier to the chamber's feedthrough and kept cable lengths as short as possible. The mirror is a protected gold coated substrate housed in a Macor mount on a plane \SI{3}{\milli\metre} above that of the centre of the quadrupole wires. More details of the laser system, optics and current driver can be found in \cite{JRushtonThesis}.

\section{Results}

\begin{figure}
\includegraphics[width=1.0\textwidth]{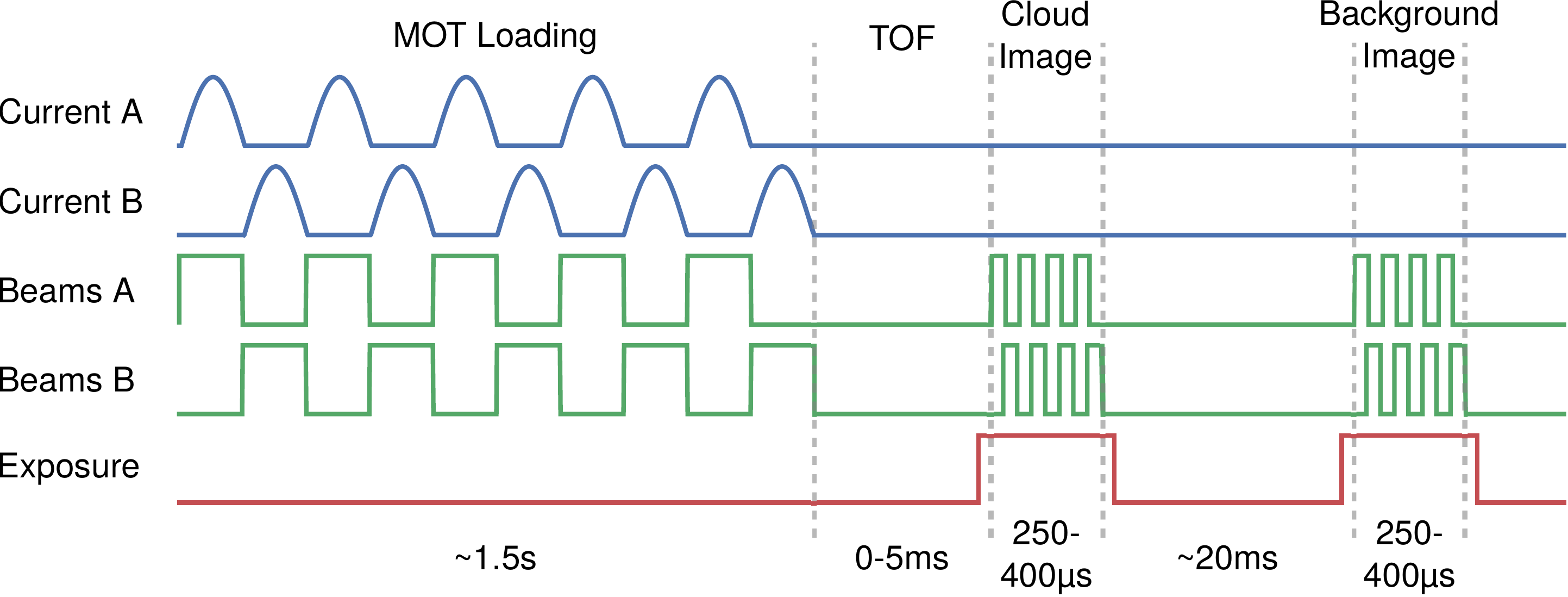}
\caption{Waveforms of the optical and current switching in the S-MOT.  The half-rectified sine waves represent the current passing through the S-MOT's wires, and is created by the circuit in figure \ref{experimental_setup}.  The digital waveforms show the shuttering of the trapping beams, the phases of which are set to synchronize with the corresponding current waveform.  This timing of the experiment is orchestrated by a Red Pitaya \protect{\cite{RedPitaya}}, an instrumentation platform containing an field-programmable gate array and a computer, which utilizes a heavily modified version of the manufacturer's arbitrary signal generator module. The duration of the stages are not to scale.}
\label{SMOT_waveform_sequence}
\end{figure}

In order to characterize the behaviour of the S-MOT the temperature and number of trapped atoms were measured with respect to the switching frequency.  Figure \ref{SMOT_waveform_sequence} illustrates the timing sequence used to perform the temperature measurements of the cold atomic ensembles.  Initially the S-MOT operates at a constant switching frequency in order to allow the trap to load. After the atom number has saturated the beams are extinguished and the current through the S-MOT's wires is then held at zero.  The atom cloud then expands ballistically for a time of flight (TOF), and \SI{25}{\micro\second} before this completes a TTL signal is sent to a ProSilica GE680 camera in order to begin the exposure of an image. The TOF ends when the beams are reactivated for an imaging pulse, which illuminates the expanded cloud so it can be photographed by the camera.  After the image has been taken the cloud is allowed to disperse and then a second image is captured to be used for background subtraction. All four beams cannot be on simultaneously during imaging due to the specific configuration of our AOM drive electronics and so the AOMs are instead switched at a consistent frequency of $\sim$\SI{30.5}{\kilo\hertz}. This frequency was chosen so that the switching period is much shorter than the duration of the imaging pulses, but it was not set to the maximum possible value in order to attain a reasonably high imaging beam intensity, which decreases as switching frequency increases. As with all of the results presented here, the data was collected in a randomized order, and so any trends cannot be attributed to the drift of any experimental parameter with respect to time.

The width of the atom cloud after each time of flight is determined by a 2D Gaussian fit to the background subtracted image.  The temperature is then determined by fitting these widths to the relation $\sigma^2=\sigma_{0}^2+(k_{\mathrm{B}}T/m)t^2$, where $\sigma_{0}$ and $\sigma$ are the Gaussian widths before and after expansion, $k_{\mathrm{B}}$ is the Boltzmann constant, $T$ is the temperature of the cloud, $m$ is the mass of ${^{85}}$Rb and $t$ is the time of flight.

The expanded clouds have a large degree of asymmetry, particularly when the S-MOT is operating at its lower switching frequencies.  The major axis of the elliptical clouds are typically aligned to the $x$ or $y$ axes of the trap, depending on the phase of the waveform when the beams are extinguished prior to the TOF.  This asymmetry is a result of a temperature difference between the $x$ and $y$ axes, due to the axes being probed at different points in their respective trapping cycles.  The temperature in both axes oscillates at the switching frequency of the trap.  This can be understood by considering that when the beams are off in a particular axis the cloud temperature along that axis heats up due to the spontaneous emission of photons, and the lack of any damping force.

As an example, if the S-MOT is in state A (illustrated in figure \ref{smot_wires_and_field}) then the atoms experience a damping force along the $x$-axis in addition to a diffusive term originating from the random nature of spontaneous emission and balanced absorption.  If the S-MOT is in state B then the RMS velocity along $x$ must be increasing, because it is only being influenced by a diffusive term from the spontaneous emission.  The temperature along the $x$ axis is thus at its minimum at the end of state A and maximum at the end of state B.  Because the two Cartesian axes are effectively $\pi$ out of phase, the maximum and minimum temperatures of the trap can be measured simultaneously.

In order to calculate the maximum and minimum temperatures in the S-MOT the system was configured to always begin the TOF at the end of state B. This ensures that the S-MOT's maximum temperatures are measured along the $x$ (hot) axis and the minimum temperatures are measured along the $y$ (cold) axis\footnote{It should be remembered that both of the trap's axes alternate between the maximum and minimum temperatures, but because we are always probing the trap in the same state for convenience we define the axes in terms of the temperature they display in our measurements.}.  To account for the asymmetry of the clouds the fitted 2D Gaussians were allowed to be elliptical and their rotation angles were unconstrained. A fixed rotation angle was not used because the alignment plays a bigger role at shorter TOFs than the temperature difference between the axes, and so in these circumstances the clouds' axes are less likely to align with those of the trap.  The fit parameters were then used to calculate the clouds' widths along the hot and cold axes, then the time dependence of these widths were used to determine the temperatures in each of the S-MOT's axes.

\begin{figure}
\includegraphics[width=1.0\textwidth]{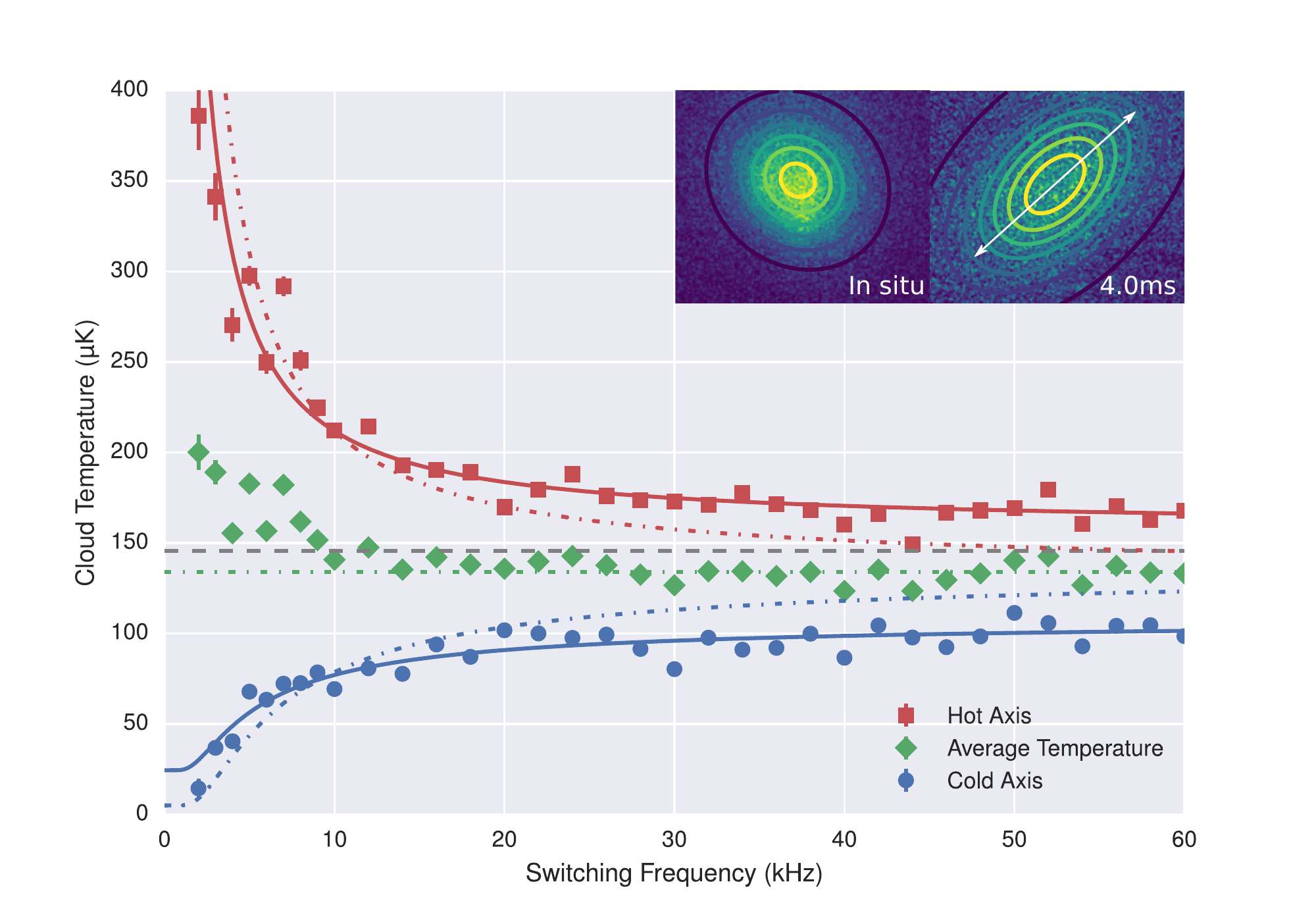}
\caption{Temperature in the hot and cold axes of the S-MOT as a function of its switching frequency at a detuning of $\delta=-2.0\Gamma$.  The dot-dashed and solid lines are fits to the data using models that are described in the text, whilst the grey dashed line indicates the Doppler temperature.  These results were taken without the use of a molasses stage and the average cooling power in each beam was \SI{2.7}{\milli\watt}. The inset shows the processed atom cloud images at a switching frequency of \SI{2}{\kilo\hertz} before and after a \SI{4.0}{\milli\second} time of flight. Eleven different flight times were used for each temperature measurement and ten images were taken for each flight time. The hot axis is rotated about \SI{48}{\degree} from the vertical due to the position of the optical column and is indicated by the white arrow.}
\label{freq_vs_temps_hot_n_cold}
\end{figure}

Figure \ref{freq_vs_temps_hot_n_cold} shows the frequency dependence of the temperatures of the hot and cold axes at a detuning of $\delta=-2.0\Gamma$. The plot also shows the averages of these temperatures, most of which are sub-Doppler and exhibit a reasonably flat frequency response. Temperatures could further be reduced with the use of an additional molasses stage. Clearly the temperature along the hot (cold) axis reduces (rises) as the switching frequency increases. The major features of this trend can be modelled by realizing that the when the trap has reached a steady state, the reduction of temperature performed in one state of the trap must equal to the increase in temperature during the other state. 

The magnitude of the momentum imparted onto an atom upon spontaneous emission is $\hbar k$, but the direction is random, uniformly distributed over the surface of a sphere.  These momentum exchanges have an average value of zero, but have a non-zero mean-squared value of $\langle p^2\rangle=\hbar^2 k^2/3$.  If the beams are directed along the axis being considered then there is additional heating due to the random absorption of the light, but also a cooling term
\begin{eqnarray}
\frac{\mathrm{d} \langle p^2\rangle}{\mathrm{d} t}=4\left(\frac{1}{3}+\frac{1}{2}\right)\hbar^2 k^2\Gamma_{\mathrm{s}} -2m\alpha v^2\,,
\end{eqnarray}
where a factor of four has been introduced to account for the effective number of beams and $\Gamma_{\mathrm{s}}$ is the scattering rate of a single beam.  The cooling term arises from the velocity dependent force, $F=-\alpha v$, which is assumed to be in the linear regime.  This force causes an exponential decay in the velocity, and hence temperature, of an atom until the minimum DC temperature, $T_{\mathrm{DC}}$ is reached
\begin{eqnarray}
T=T_0\exp\left(-\frac{2\alpha}{m}t\right) + T_{\mathrm{DC}}\,,
\end{eqnarray}
where $T_0 + T_{\mathrm{DC}}$ is the initial temperature and the minimum DC temperature is allowed to be sub-Doppler.  If the S-MOT is operating at a frequency $f$ then the atoms are not permitted to reach $T_{\mathrm{DC}}$ and instead cool for a duration of $t=1/2f$ until reaching a temperature of $T_{\mathrm{C}}$.  The S-MOT is in a steady state when the degree of cooling over a half period equals the heating over the next half period
\begin{figure}
\includegraphics[width=1.0\textwidth]{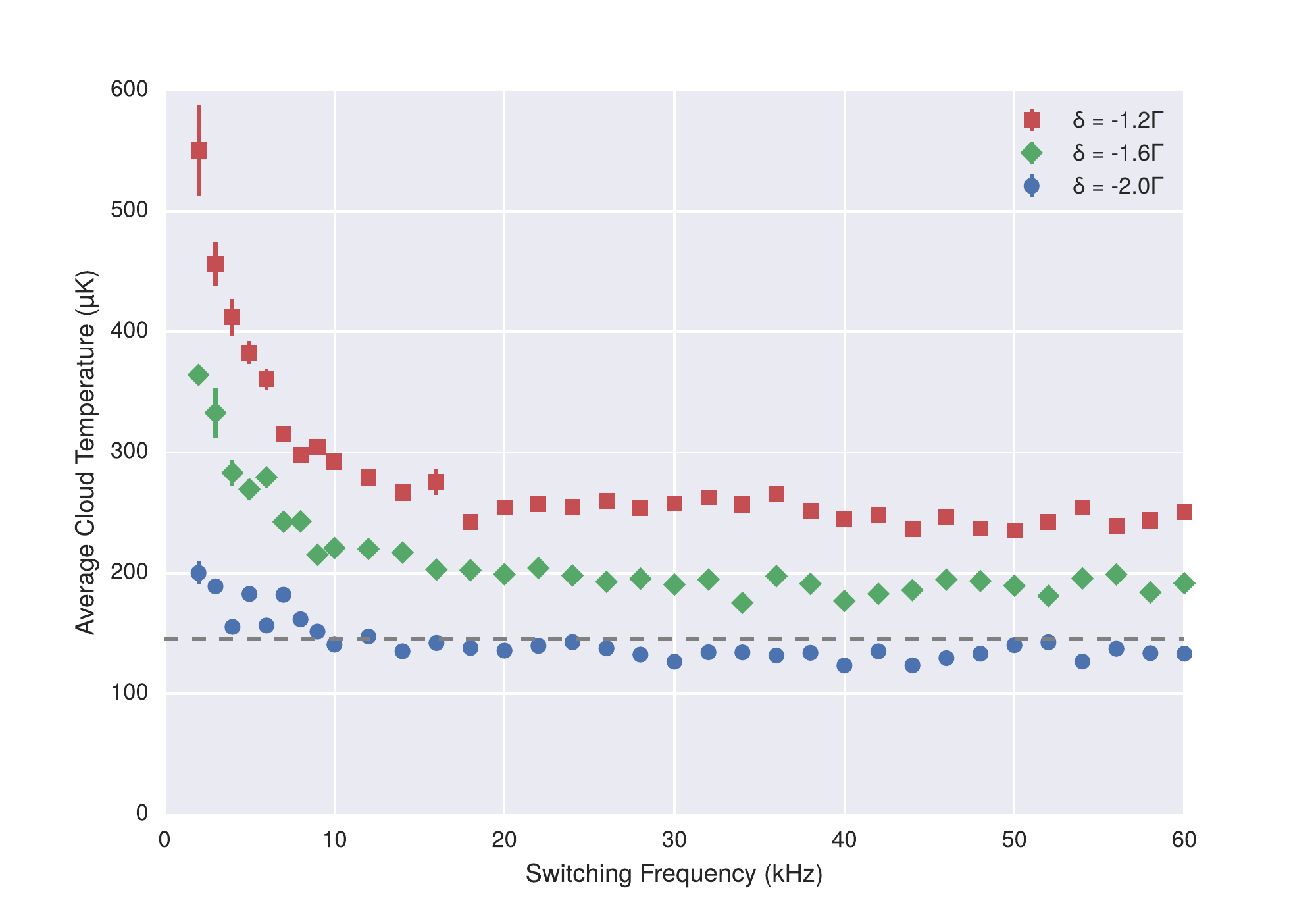}
\caption{Average cloud temperature in the S-MOT as a function of its switching frequency.  The average temperatures are found by taking the mean of the temperatures of the hot and cold axes, as shown in figure \ref{freq_vs_temps_hot_n_cold}.  These results were taken without the use of a molasses stage and the grey dashed line indicates the Doppler temperature. The average cooling power in each beam was \SI{2.7}{\milli\watt}.}
\label{freq_vs_av_temps}
\end{figure}
\begin{eqnarray}
\int_{0}^{\frac{1}{2f}} \frac{10}{3}\hbar^2 k^2\Gamma_{\mathrm{s}}\,\mathrm{d}t - \int_{0}^{\frac{1}{2f}} 2m\alpha v^2\,\mathrm{d}t&=-\int_{\frac{1}{2f}}^{\frac{1}{f}} \frac{4}{3}\hbar^2 k^2\Gamma_{\mathrm{s}}\,\mathrm{d}t\,,
\end{eqnarray}
from which $T_0$ can be found to be
\begin{eqnarray}
T_0=\frac{1}{mf}\frac{\frac{7}{3}\hbar^2 k^2\Gamma_{\mathrm{s}}k^{-1}_{\mathrm{B}} - T_{\mathrm{DC}}\alpha}{\left[1-\exp\left(-\frac{\alpha}{mf}\right)\right]}\,,
\end{eqnarray}
where $k_{\mathrm{B}}$ is the Boltzmann constant.  The hot temperature is given by $T_{\mathrm{H}}=T_0 + T_{\mathrm{DC}}$, and the cold temperature is given by $T_{\mathrm{C}}=T_0\exp\left(-\alpha/mf\right) + T_{\mathrm{DC}}$.  In the limit of infinite frequency both of these temperatures approach the asymptote
\begin{eqnarray}
T_{f\rightarrow\infty}&=\frac{7}{3}\frac{\hbar^2 k^2\Gamma_{\mathrm{s}}}{k_{\mathrm{B}}\alpha}\,.
\end{eqnarray}
This model's values of $T_{\mathrm{H}}$, $T_{\mathrm{C}}$ and $T_{f\rightarrow\infty}$ have been fit to the data in figure \ref{freq_vs_temps_hot_n_cold} and are shown as a series of dot-dashed curves. The general trend for the temperatures of the hot and cold axes is reflected in the data, however there is a divergence at higher frequencies. This discrepancy may be due to the model being too simplistic, but could also be partially caused by imperfect switching of the beams due to the non-negligible rise and fall times of the AOMs. This effect can be incorporated into our expressions of $T_{\mathrm{H}}$ and $T_{\mathrm{C}}$ by the transformation $f\Rightarrow f'=f/(1+2f\Delta t)$, where $\Delta t$ is the minimum on-time of the beams, limited by the frequency independent fall time.  The solid lines of figure \ref{freq_vs_temps_hot_n_cold} shows a fit of the model using this transformation, but it should be realized that the fit parameter for $\Delta t$ cannot be fully accounted for by the apparatus as its value is unrealistically large (\SI{30}{\micro\second}). Our fitting finds a minimum DC temperature of $T_{\mathrm{DC}}\sim$ \SI{24}{\micro\kelvin} which is a reasonable lower bound for a rubidium MOT incorporating sub-Doppler cooling mechanisms. 

\begin{figure}
\includegraphics[width=1.0\textwidth]{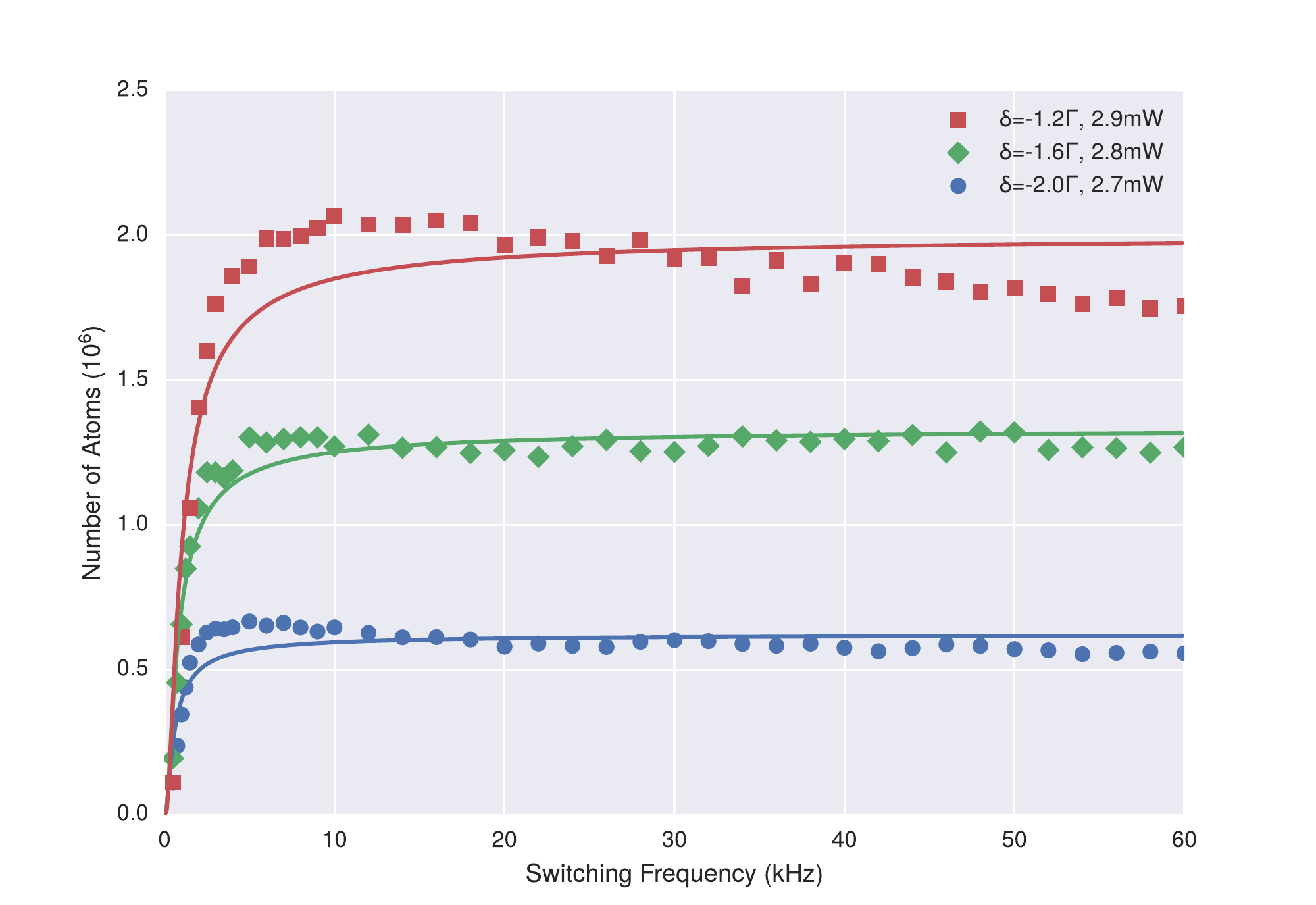}
\caption{Atom number in the S-MOT as a function of its switching frequency. The solid curves are fits to the data using the model described by equation \ref{atom_number_scaling_model}.}
\label{freq_v_N}
\end{figure}

We have also investigated the effect of switching on the average temperature and atom number for various detunings of the cooling laser, as shown in figures \ref{freq_vs_av_temps} and \ref{freq_v_N}. As found in most MOTs \cite{PLettSubDoppler}, lower temperatures are reached with increasing red-detuning due to the greater influence of sub-Doppler cooling mechanisms. The atom number, on the other hand, was observed to rapidly increase with switching frequency below $\sim$\SI{10}{\kilo\hertz}.  This can be understood by considering the time it takes for an atom to traverse the overlap region of the trapping beams.  Take the extreme example of an atom moving in the positive $x$ axis while the S-MOT is in state B. If the switching frequency is low enough such that the atom can pass through the trapping region before the state of the trap changes, then the atom clearly cannot be captured as it does not experience any decelerating force.

As in the previous model, we assume that there exists a velocity dependent force in the S-MOT of $F=-\alpha v$ leading to an exponential deceleration of atoms moving in the trapping region.  This deceleration produces a stopping distance which has to be less than the beam diameter, $L$, in order for them to be trapped within a standard MOT.  This results in a capture velocity of $v_c=L\alpha/m$, however the S-MOT only exerts a velocity dependent force in each axis for half of the time that an atom takes to traverse the overlap region, and so the effective beam width is halved.  This leads to the calculation of the effective capture velocity in the S-MOT, half the conventional capture velocity
\begin{eqnarray}
v'_c=\frac{v_c}{2}=\frac{L\alpha}{2m}\,.
\label{capv}
\end{eqnarray}
At high frequencies it is valid to treat the maximum effective stopping distance as equal to half the beam diameter, but at lower frequencies the distance over which the atoms decelerate will depend strongly both on the switching frequency as well its phase.

The worst case scenario for an atom with initial velocity $\vec{v}=v_{0}\hat{x}$ undergoing trapping is to enter the beam overlap region just as the S-MOT begins state B. In this case the atom must wait until the time $t=1/2f$ before it starts decelerating, during which it moves a distance $v_{0}/2f$.  Throughout the second half of the first period,  $1/2f<t<1/f$, the then active beam pair in the $xz$-plane acts to provide a force proportional to the atom's velocity, exponentially slowing it to $v_{1}=v_{0}\exp(-\alpha /2mf)$. During the period $1/f< t< 3/2f$ the atom once again does not experience a decelerating force, and so travels a distance of $v_{1}/2f=v_{0}\exp(-\alpha /2mf)/2f$.  In general, by time $t=n/f$ the atom slows to
\begin{eqnarray}
v_{n}=v_{0}\exp\left(-\frac{n\alpha}{2mf}\right),
\end{eqnarray}
so during the period $n/f<t<(2n+1)/2f$ it moves a distance
\begin{eqnarray}
\frac{1}{2f}v_{n}=\frac{1}{2f}v_{0}\exp\left(-\frac{n\alpha}{2mf}\right).
\end{eqnarray}
The total distance travelled during trapping can be found by introducing an infinite summation, the value of which must be less than or equal to the beam diameter, $L$, in order for the atom to be captured:
\begin{eqnarray}
\label{stopdistline1}
L&\ge\frac{m}{\alpha}v_{0} + \frac{1}{2f}v_{0} + \sum_{n=1}^{\infty}\frac{1}{2f}v_{0}\exp\left(-\frac{n\alpha}{2mf}\right)\\
&\ge\frac{m}{\alpha}v_{0} + \frac{1}{2f}v_{0}\frac{1}{1 - \exp\left(-\frac{\alpha}{2mf}\right)}\,,
\label{stopdistline2}
\end{eqnarray}
where the first term on the right hand side is the distance travelled while decelerating, i.e. the standard MOT stopping distance, and the $v_0/2f$ term is the initial distance travelled before deceleration first begins.  This second term is not present if the atom enters the trapping region at the optimum phase of the waveform.  The total of the right hand side has the limit of $2mv_0/\alpha$ as the frequency approaches infinity. When this limit is equated to the maximum stopping distance, $L$, it predicts the effective capture velocity $v'_{c}$ as expected.  Rearranging this equation allows us to find the minimum frequency required to capture atoms with initial velocity $v_{0}<v'_c$ 
\begin{eqnarray}
\label{minimum_frequency}
f_{\mathrm{min}}= \frac{\alpha}{2m} \left\{W\left[\left(1-\frac{L\alpha}{mv_{0}}\right)\exp\left(1-\frac{L\alpha}{mv_{0}}\right)\right] + \frac{L\alpha}{mv_{0}} - 1\right\}^{-1}\,,
\end{eqnarray}
where $W[x]$ is the Lambert W function. The atom number in a MOT is proportional to $v^4_c$ \cite{Lindquist1992}, so by rearranging again and assuming the switching frequency does not influence the S-MOT's loss rate we would expect the relative atom number to scale as
\begin{eqnarray}
\label{atom_number_scaling_model}
\frac{N}{N_{\mathrm{max}}}\propto\left[\frac{v_0}{v'_c}\right]^4=16\left[\frac{1-\exp\left(-\frac{\alpha}{2mf}\right)}{\frac{\alpha}{2mf} + 1 - \exp\left( -\frac{\alpha}{2mf} \right)}\right]^4\,,
\end{eqnarray}
where $N_{\mathrm{max}}$ is the maximum atom number obtained at infinite switching frequency.  This relationship has been fit to the data in figure \ref{freq_v_N} and well describes its dominant features.  Observing the atom number data we can see that the minimum switching frequency seems to be dependent on the detuning of the cooling beam.  This is apparent by the shift of the position of the `elbow' of each curve. This should be expected because a smaller detuning results in a larger value of the damping coefficient, $\alpha$, and equation \ref{atom_number_scaling_model} shows that for a fixed switching frequency, higher values of $\alpha$ result in the S-MOT capturing a smaller proportion of its maximal atom number.

Figure \ref{freq_v_N} also shows a trend of decreasing atom number as the switching frequency is raised above \SI{10}{\kilo\hertz}.  This behaviour is not present in either our theoretical model or our more detailed Monte Carlo simulations, and it was thought that the frequency dependence was impressed by an imperfect RF switch, which is used to direct power between the switching AOMs.  Replacing the switch with a different model has helped to reduce the drop off, however the effect is still present.  We suspect it can be attributed to systematic error, perhaps due to a slight misalignment of the trap, or due to the frequency dependence of the beam power which may not have been accounted for completely.

\section{Discussions}
\label{Discussions}
Our results have shown that the S-MOT achieves characteristics similar to conventional mirror-MOTs, albeit with a slightly reduced atom number due to the switching nature of the trap. An optimum switching frequency of approximately \SI{20}{\kilo\hertz} results in the greatest atomic density of \SI{3e09}{\per\centi\metre\cubed}, with very little variation about this value. The asymmetric temperatures of the atom clouds produced at low trapping frequencies allows one to prepare samples of atoms that are very cold along one axis. The limited optical access of the system prevented us from determining the temperatures normal to the mirror, however we do know that this axis undergoes continuous (DC) cooling and so we would expect it to be the coldest of them all. We found that the S-MOT is very sensitive to the alignment of its beams, so great care must be taken to prevent interference fringes disturbing the clouds, destroying their rotational symmetry and limiting their atom number.  

Due to the geometry of the design the edges of the trapping beams get clipped by the S-MOT's wires.  The maximum $1/e^2$ beam diameter is hence limited to around \SI{7.5}{\milli\metre} and in turn defines the trap's maximum possible capture volume. This restricts the atom number in the S-MOT, which is lower than that of a conventional M-MOT due to the maximum effective stopping distance being half of a DC MOT with the same beam size. This beam geometry does have its advantages though, in particular their \SI{45}{\degree} angle of incidence results in a lower amount of scatter than traps which have their beams normal to their surfaces, and allows for high NA collection optics to be placed closer to the atom cloud.

In earlier versions of this experiment we drove the half-wave rectifiers of the S-MOT with commercially available audio amplifiers, but found them susceptible to overheating and unable to deliver their rated power. In contrast, our bespoke current driver can deliver a stable \SI{40}{\ampere\peakpeak} sine wave through the S-MOT's rectifying circuitry (figure \ref{experimental_setup}) at frequencies up to \SI{60}{\kilo\hertz}. This device can deliver any waveform to the S-MOT's wires, providing it is kept within its safe operating area, and this property is of particular use as it can be employed to rapidly extinguish the trap's magnetic fields.  Here we explored the behaviour of the S-MOT at a range of switching frequencies, but in general multi-frequency operation may not be required.  If single-frequency mode is sufficient then a resonant tank circuit could be used to drive the S-MOT's wires, possibly incorporating a transformer, allowing for the electronics to be greatly simplified and the power dissipation to be vastly reduced.

A peculiarity of the S-MOT is the restricted movement of its atom clouds upon applying an external bias field.  If a magnetic field bias is applied in the $z$ direction then this will not move the cloud in the $z$ axis, but along the line $y=-x$.  To bring the cloud closer to the mirror's surface a field needs to be applied along $(\hat{y}-\hat{x})/\sqrt{2}$, however this will result in the cloud's $z$ position oscillating throughout the switching period. This is due to the sinusoidal current passing through the S-MOT's wires combining with the bias to create a time dependence of the position of the magnetic field zero. This may have produced a small amount of heating and possibly the unexpected drop off in atom number with increasing frequency in the data-sets presented.  These oscillations are not present if square current pulses pass through the S-MOT's wires instead of half-rectified sine waves, however these waveforms are more likely to endanger the amplifier and cause electromagnetic interference, which is the reason this scheme was abandoned earlier in the experiment.  By granting individual control of alternating and direct currents passing through each of the S-MOT's wires it would be possible, in future versions of the trap, to provide background field cancellation without additional external nulling coils. This same scheme would also enable movement of the magnetic field zero to any point in the $xy$-plane and, additionally, the wires could be implemented in a dual-plane design in order to give control of the $z$ position of the magnetic field zero. This would not only enable manipulation of the cloud's location in the trap without the aforementioned oscillations, it would allow for a bias field to be applied in any direction.

We have successfully demonstrated a new, dynamic, design of mirror-MOT which is suitable for use with microfabricated atom chips and any vacuum chambers with restricted optical access.  The design itself is amenable to microfabrication due to the absence of out of plane wires or coils and does not use any frequency selective optics, permitting it to trap different atomic species simultaneously. 

\ack
This work was supported by funding from RAEng, EPSRC, and the UK Quantum Technology Hub for Sensors and Metrology under grant EP/M013294/1.  The authors would like to thank Mark Bampton for machining work including the S-MOT's ceramic structure, Gareth Savage for frequent advice regarding electronics and Eric J Boere from Apex Microtechnology for assistance in designing the current amplifier.  We would also like to thank Aiden Arnold and Tim Freegarde.

\section*{References}

\bibliographystyle{iopart-num_w_doi}
\bibliography{SMOT}
\end{document}